    \documentstyle[aps,prl,amstex,twocolumn,psfig]{revtex} 

\begin{document} 
 
\draft 
\title{Crossover from a square to a hexagonal pattern\\ 
in Faraday surface waves} 
\author{C.~Wagner$^1$, H.~W.~M\"uller$^{2}$, and K.~Knorr$^1$} 
\address{$^1$ Institut  f\"ur Technische Physik, Universit\"at 
des Saarlandes Postfach 151150, D-66041 Saarbr\"ucken, Germany\\
$^2$Max Planck Institut f\"ur Polymerforschung, Ackermannweg 10,
D-55128 Mainz, Germany 
%$^3$Institut f\"ur Theoretische Physik, Universit\"at 
%des Saarlandes Postfach 151150, D-66041 Saarbr\"ucken, Germany
} 
\maketitle 
 
\begin{abstract} 
We report on surface wave pattern formation in a Faraday experiment operated at
a very shallow filling level, where modes with a subharmonic and 
harmonic time dependence interact. Associated with this distinct temporal
behavior are different pattern selection mechanisms, favoring squares or hexagons, 
respectively.
In a series of bifurcations running through a pair of superlattices 
the surface wave pattern transforms between the two
incompatible symmetries. The close analogy to 2D and 3D crystallography is pointed out. 
\end{abstract} 
% 
% 47.50.+d Non-Newtonian fluid flows
% 47.54.+r  Pattern selection; pattern formation
% 47.20.-k : Hydrodynamic stability 
% 47.10.+g : General theory (of fluid dynamics) 
% 47.20.Ky : Nonlinearity (including bifurcation theory) 
% 47.20.Ma Interfacial instability 
% 47.20.Lz Secondary instability
% 47.20.Gv Viscous instability 
% 47.15.Cb Laminar boundary layers 
%new categories in 1999 //05 is Statistical physics and thermodynamics
%                       //45 is Classical mechanics of discrete systems
% 05.45.-a :Nonlinear dynamics and nonlinear dynamical systems (see also 45 Classical mechanics of discrete systems)  
%45.70.Qj Pattern formation
    %changes1999
%**************************************************************** 
%*                                                              * 
%*      begin of text                                           * 
%*                                                              * 
%**************************************************************** 
\narrowtext  
When a fluid layer is vibrated vertically   patterns of standing waves occur
at the liquid-air interface. This experiment, first studied by
Faraday  \cite{faraday31} in 1831, has become a paradigm for the 
investigation of spontaneous pattern formation \cite{miles90}. From an experimental 
point of view, 
the Faraday setup is particularly attractive since the characteristic length and time
 scales are under 
external control and relaxation times are pretty short.
If the 
vibration signal is sinusoidal and weakly supercritical one generically
observes well 
ordered standing wave patterns in the form of squares or lines, oscillating with
half the frequency of the excitation (subharmonic response). By carefully tuning the 
drive frequency within the
crossover regime between gravity and capillary surface waves, more complicated 
(quasi-periodic) patterns with an 8 or even 10-fold point symmetry 
have been predicted and observed \cite{chen97}.  Edwards and Fauve \cite{edwards94} 
introduced the idea of a two-frequency excitation signal.
The resulting  interplay
between the competing modes gave rise  to unexpected new phenomena like hexagonal, 
triangular, quasi-periodic, spatially localized structures and even 
superlattice type patterns \cite{kudrolli98,arbell98}. However, the large number of 
control parameters  of a multi-frequency drive signal
renders such an experiment problematic as far as a systematic exploration of parameter space
is concerned and it lacks an intuitive understanding of the fundamental principles. 
By use of a viscoelastic liquid, it has been shown \cite{wagner99}
that the familiar subharmonic Faraday resonance can be preempted by a 
synchronous (harmonic) response, leading to a bicriticality as well.
However, such fluids are more complicated than Newtonian liquids and the 
characterization of their nonlinear rheology is incomplete.
For Newtonian fluids the parameter region in which the response switches 
from subharmonic to harmonic is difficult to access 
\cite{benjamin54,kumar96,cerda97,muller97} since it requires a very
shallow filling depth, large shaking elevations and thus a powerful vibrator.
This article describes a systematic investigation of a Newtonian liquid under a single-frequency
drive operated close to the bicritical crossover. The interesting new feature 
is a ''phase transition'' from a
quadratic pattern to a structure of hexagonal symmetry. The symmetry change 
proceeds via a sequence of superlattices with 2-fold and 6-fold point symmetry.     

Our new shaker system has a force of 4800 N and a maximum  
peak-peak elevation of 5.4 cm. It is operated at a rather low drive frequency of $8-10 Hz$.
 The combination of these parameters with a filling depth of $0.7mm$, only, is necessary to
  access the harmonic-subharmonic bicriticality. 
The container with an inner diameter of $290mm$ (equivalent to about $15$ times the
wavelengths) is sealed by
a glass plate and temperature controlled at $25^\circ  \pm 0.1^\circ $C.
At this temperature the sample fluid (low viscosity  Silicon oil, Dow Corning 200) is
specified by a kinematic viscosity $\nu = 9.7\times 10^{-6}m^2/s$, 
a surface tension of $\sigma = 0.0201 N/m$ and a 
density $\rho = 934 kg/m^3$. 
The actual acceleration 
experienced by the container is recorded 
by a piezoelectric device. A feedback loop control limits 
disturbing anharmonicities to a level of less than 0.2\%.
Our visualization method for the evaluation of pattern symmetries has been
described elsewhere \cite{wagner99}.  Although this simple light reflection technique
provides high contrast pictures, the relation between the recorded intensity and 
the surface profile is not trivial. Nevertheless for simple structures like squares, 
we were able to reconstruct the profile by the following procedure: 
Starting from an estimated surface profile composed of a 
small number of spatial Fourier modes, we computed the light distribution of the 
expected video image by means of a ray tracing algorithm. 
Then we adapted the mode amplitudes and their relative phases to optimize
the agreement between the calculated and recorded video pictures. 
 
\begin{figure}
\psfig{file=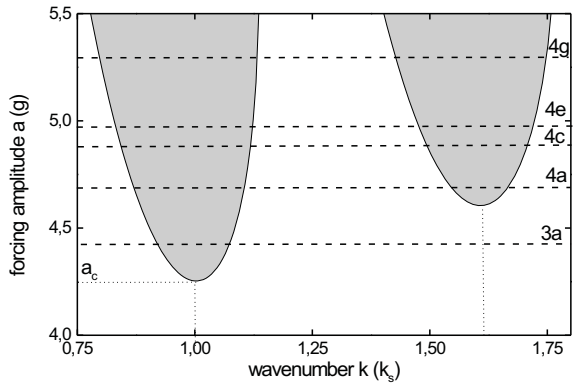,width=1.\columnwidth,angle=0} 
\caption[] 
{ 
Neutral stability diagram for an excitation frequency $\Omega=2\pi\times 9.5Hz$ slightly above
the bicritical point $\Omega_B=2 \pi \times 8.7Hz$. Within the left (right) 
resonance tongue plane 
wave perturbations (wavenumber k) with a subharmonic (harmonic) time dependence become
unstable. The absolute minimum ($a_c=4.25 g$, $k_s=423 m^{-1}$) determines the 
threshold for the onset
of Faraday waves. %The abscissa is measured in units of the critical wavenumber $k_s$. 
The dashed horizontal lines indicate the drive amplitudes at which the patterns of Figs.3
and 4 are taken.
} 
\label{fig1} 
\end{figure}

\begin{figure} 
\psfig{file=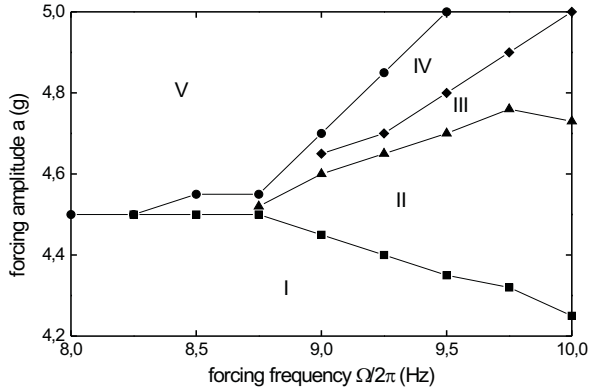,width=1.\columnwidth,angle=0} 
\caption[] 
{ 
Phase diagram of the observed 
nonlinear patterns. Region $I$: flat surface;  
$II$: subharmonically oscillating squares as depicted in Fig.~\ref{fig3}a,b. 
$III$: $\sqrt[]{2} \times
 \sqrt[]{2}$ superlattice (Fig.~\ref{fig4}a,b); 
 $IV$: $\sqrt[]{3} \times  \sqrt[]{3}$ superlattice (Fig.~\ref{fig4}e,f);
$V$: harmonically oscillating hexagons (Fig.~\ref{fig4}g,h). The symbols mark the
experimental observed transition points between the phases.  
 
} 
\label{fig2} 
\end{figure}

The linear stability theory \cite{kumar94} evaluated for our fluid predicts that the bicritical
 threshold, at which the system changes from the harmonic instability at lower to the subharmonic
  instability at higher frequencies, occurs at a drive frequency of $\Omega_B/2 \pi=8.7Hz$. This is in
   good agreement with the present results. Fig.~\ref{fig1} depicts a calculated neutral stability diagram 
(acceleration amplitude $a$ vs. wavenumber $k$) for an excitation frequency 
$\Omega/2 \pi=9.5 Hz$ slightly above $\Omega_B$. The minimum of the 
left (right) resonance tongue defines the wavenumber $k_s$ ($k_h$) 
of the subharmonic S (harmonic H) instability. 
The absolute minimum $a_c$ corresponds to the onset of Faraday waves.
 Fig.~\ref{fig2} presents a phase diagram classifying the  
surface patterns as observed during a quasistatic amplitude ramp at fixed drive
 frequencies. Starting from a subcritical drive $\epsilon=a/a_c-1=-2\%$ 
the amplitude $a$ is increased by steps of $0.2\%$.
After each increment the scan is suspended for 240 sec and a surface picture is 
taken thereafter. Having reached $\epsilon=10\%$ 
the ramp is reversed. There is no noticeable hysteresis for the primary onset between upward
and downward scans.

Within the drive frequency region of the harmonic Faraday instability 
($\Omega<\Omega_B$) the bifurcation scenario is rather conventional: Entering 
subregion $V$ from below (see Fig.~\ref{fig2}), we find a perfect hexagonal surface tiling
 which persists up to  the maximum drive amplitude. 

\begin{figure} 
\psfig{file=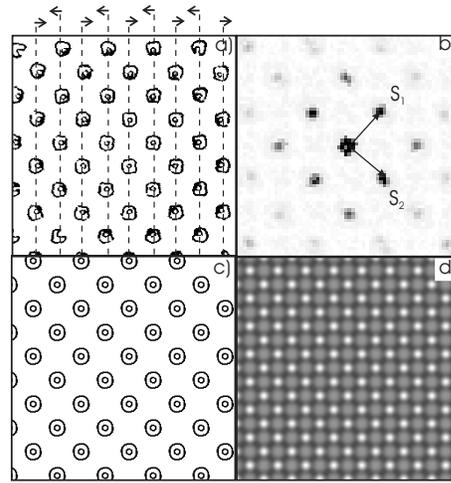,width=6cm,angle=0} 
\caption[] 
{ 
a,b): Photograph and Fourier spectrum  of the subharmonic square pattern
(region II in Fig.~\ref{fig1}).
All video images show the central region of the cylindrical container, whose diameter is
about twice as large. c) video image computed by the ray tracing technique
d) density plot of the associated 2D surface profile.

% d): Density plot of the surface profile as obtained from 
%our ray tracing technique. Best agreement between calculated video image (c) and 
%experiment has been achieved for the following modes amplitudes and phases: 
%$A_{S1}=A_{S2}=1$, $A_{S1 \pm S2}=1.5$, $A_{2 S1}=A_{2 S2}=0.5$,$A_{2 (S1 \pm S2)}=0.5$,$\phi_i=0$. 
} 
\label{fig3} 
\end{figure}

The focus of the present paper is on the subharmonic region $\Omega>\Omega_B$, where
the onset pattern  (region $II$) is quadratic, while hexagons (region $V$) occur
at rather elevated shaking amplitudes as a higher bifurcation. 
Our aim is to describe the bifurcation sequence, which results from an amplitude scan 
at the fixed drive frequency of $\Omega=9.5 Hz$ (gravity wave regime). 
The primary Faraday pattern (region $II$ in Fig.~\ref{fig2}) has a 
subharmonic time dependence and exhibits a perfect quadratic symmetry as shown 
in Fig.~\ref{fig3}a. The associated 
spatial power spectrum (Fig.~\ref{fig3}b) indicates  the fundamental
wave vectors ${\bf k}_{S1}$ and ${\bf k}_{S2}$ but also pronounced contributions from
higher harmonics, in particular ${\bf k}_{S1}+{\bf k}_{S2}$. The appearance of square 
patterns in low viscosity gravity waves agrees with a small amplitude theory
expanded around the subharmonic instability threshold \cite{chen97}. 
At those small values of  $\varepsilon$ the competing harmonic Faraday modes 
do not noticeably affect the pattern selection process. 
The shallow filling level used in our setup makes the surface elevation profile
very anharmonic with high, narrow tips but broad, shallow hollows.
We have re-constructed the 
surface profile belonging to Fig.~\ref{fig3}a by our ray tracing algorithm. 
The spatial dependence of the interface deformation is decomposed according to
\begin{equation}
\eta({\bf r})=\sum_i A_i \cos{({\bf k}_i \cdot {\bf r} + \phi_i)} ,
\label{super}
\end{equation}
where ${\bf k}_i \in 
\{ {\bf k}_{S1},\, {\bf k}_{S2},\, ({\bf k}_{S1} \pm {\bf k}_{S2}),\,
2 {\bf k}_{S1},\, 2{\bf k}_{S2},\, 2({\bf k}_{S1} \pm {\bf k}_{S2})\}$ and
${\bf r}=(x,y)$.
A density plot with the gray level proportional to the 
local surface elevation is given in Fig.~\ref{fig3}d together with the related video image (c) 
as computed by ray tracing the reflected light.

\begin{figure} 
\psfig{file=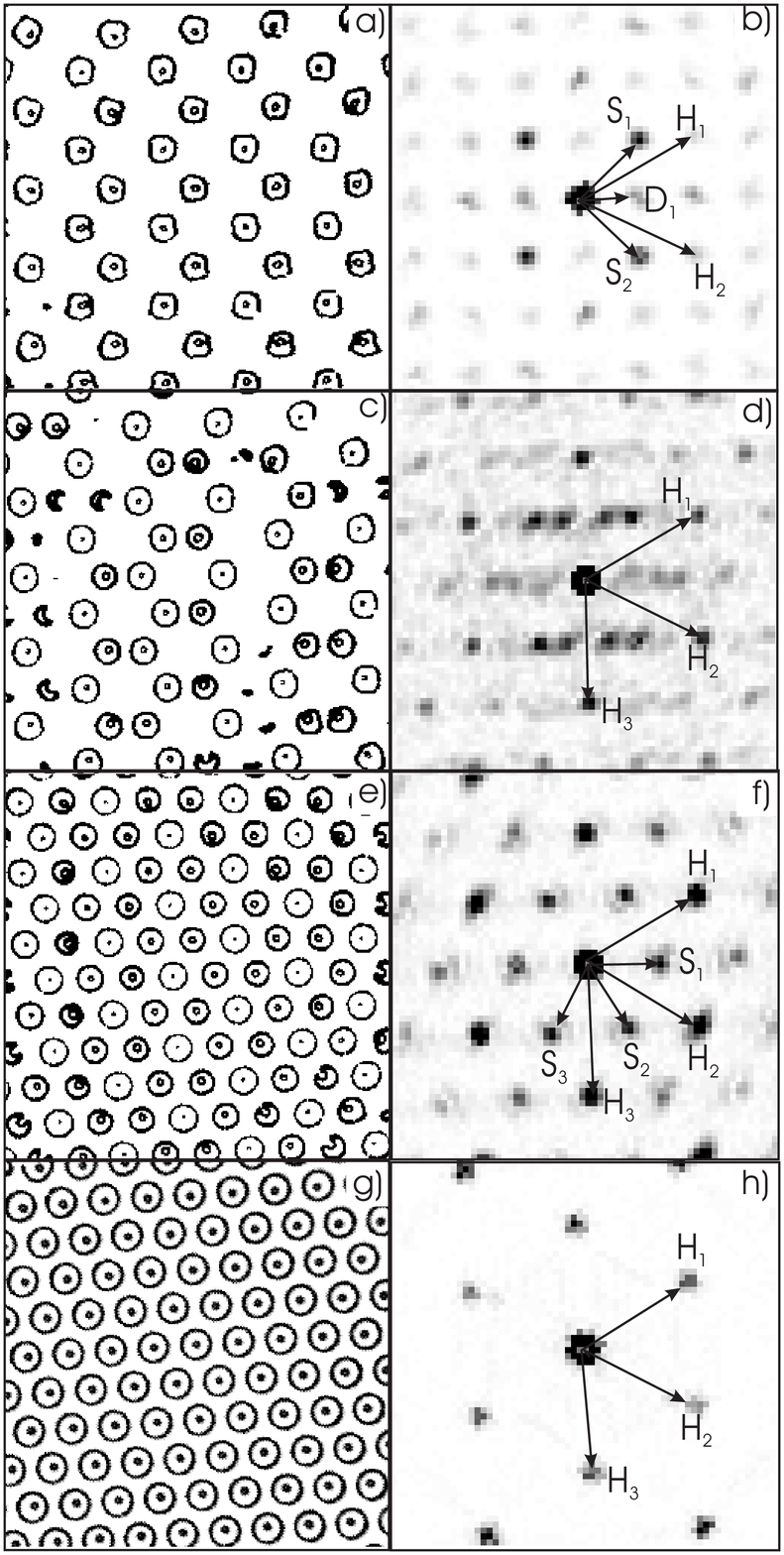,width=6cm,angle=0} 
\caption[] 
{ 
Photographs and related Fourier spectra taken at drive amplitudes as indicated in Fig.~1;
a,b): $\sqrt{2} \times \sqrt{2}$ superlattice corresponding to region III of Fig.~\ref{fig2};
c,d): ''quasi-hexagonal'' transient at the cross-over between $III$ and $IV$;
e,f): $\sqrt[]{3} \times \sqrt[]{3}$ superlattice  (region IV); g,h) hexagonal pattern 
(region $V$). 
} 
\label{fig4} 
\end{figure}

With increasing drive amplitude the harmonic Faraday instability gradually gains influence
 upon the pattern selection dynamics. When entering region $III$ we observe a 
continuous transition to a $\sqrt{2}\times\sqrt{2}$
superlattice (see Fig.~\ref{fig4}a,b), characterized by a new subharmonic mode with
wave vector ${\bf k}_{D1}$. 
Even though the associated Fourier spectrum also shows a second mode ${\bf k}_{D2}$ 
 orthogonal to the first one, the amplitude ratio $A_{D1}/A_{D2}\simeq 4$ 
indicates that ${\bf k}_{D1}$ is strongly prevailing. Therefore the original
 quadratic invariance of the pattern is broken and replaced by the simpler 
rectangular symmetry. The surface profile as derived by ray tracing 
suggests that the mode ${\bf k}_{D1}$ enters 
Eq.~\ref{super}
 in the form $\cos{({\bf k}_{D1} \cdot {\bf r}-\pi/2})$. Assuming $\phi_{S1}=\phi_{S2}=0$ 
(by a proper choice of the origin) the cubic nonlinearity $A_{S1} A_{S2} A_{D1}^*$ 
enters the amplitude equation for $A_{D1}$ and thus
restricts the relative phase $\phi_{D1}$ to  $0$ or $\pi/2$,
depending on the sign of the related coupling coefficient. The former (latter) value
is associated with a modulative (displacive) 
mode. Details will be published elsewhere. 
For our system the second case applies, which becomes evident by comparing the video images 
Figs.~\ref{fig3}a and
\ref{fig4}a: With increasing order parameter $A_{D1}$ the rows of
elevation maxima (connected by dashed lines in Fig.~\ref{fig3}a) are displaced in 
opposite directions as indicated by the arrows. 
The displacive mode ${\bf k}_{D1}$ is excited by a pair of wavevector triads 
according to the geometrical relations 
${\bf k}_{D1}={\bf k}_{H1}-{\bf k}_{S1}$ and ${\bf k}_{D1}=
{\bf k}_{H2}-{\bf k}_{S2}$. The modes ${\bf k}_{H1}$, ${\bf k}_{H2}$ oscillate
synchronously with the drive. Their contribution to the triad-couplings becomes
energetically efficient since $|{\bf k}_{H1}|=
|{\bf k}_{H_2}|=\sqrt{5/2}\simeq1.58$ coincides almost exactly with the  wavenumber
$k_h\simeq 1.59$ associated with the harmonic Faraday instability (Fig.1). 

The transition from region $III$ to $IV$ is accompanied by a very slow 
rearrangement of the pattern and the occurrence of defects. 
A snapshot taken  during this crossover process is shown in
Fig.~\ref{fig4}c. As $\varepsilon$ is increased, the displacive mode ${\bf k}_{D1}$ 
gradually dies out and thus  ${\bf k}_{H1}$ and ${\bf k}_{H2}$ enter into a new 
triad together with the nonlinear harmonic mode 
${\bf k}_{H3}={\bf k}_{S2}-{\bf k}_{S1}$. Even though 
$|{\bf k}_{H3}|=\sqrt{2}$ does not exactly match the unstable wavenumber band 
around $k_h$ (see Fig.~\ref{fig1}) the drive amplitude is apparently 
high enough to allow this detuning. The six-armed star of wavevectors 
${\bf k}_{Hi}$ shown in Fig.~\ref{fig4}d already suggests a 6-fold rotational invariance, 
but the true symmetry of this ''quasi-hexagonal''
transient state is still  2-fold:  The angles between ${\bf k}_{H1}$ and 
${\bf k}_{H2}$ and between ${\bf k}_{H2}$ and ${\bf k}_{H3}$ are  
$56^\circ$  and $62 ^\circ$, respectively, rather than $60 ^\circ$. 
>From the horizontal streaks of the 
Fourier spectrum of Fig.~\ref{fig4}d it can be seen that the translational coherence 
in $x$-direction is in a process of disintegration, while the spatial periodicity 
in $y$-direction and the 2-fold
 point symmetry are still preserved.  Nevertheless the mutual resonance among
the ${\bf k}_{Hi}$ is the  precursor of the hexagonal symmetry, which will follow.

The transient  re-orientation process comes to a halt when the pattern has 
accomplished the ideal hexagonal symmetry (region IV and Figs.~\ref{fig4}e,f).
At this stage the harmonic Faraday 
modes govern the  pattern selection process. The structure depicted in 
 Figs.~\ref{fig4}e,f can be understood
 as a  $\sqrt{3}\times \sqrt{3}$ superlattice of the hexagonal lattice of the final state (region $V$).
  Apparently, the 
subharmonic contributions to the Fourier spectrum are dynamically slaved, since the 
6-fold symmetry dictated by the ${\bf k}_{H}$-modes is recovered in number as well
as in orientation by the new subharmonic set   $\{ 
{\bf k}_{S1}, \, {\bf k}_{S2}, \, {\bf k}_{S3} \}$.
Note that the subharmonic time dependence does not allow a mutual resonance among the 
${\bf k}_{Si}$. The  relative $30^\circ$-orientation
between the S-star and the H-star  results from the 
triad cross-coupling between harmonic
and subharmonic modes. The appropriate geometrical resonance conditions
${\bf k}_{S1}+ {\bf k}_{S2}={\bf k}_{H1}$, etc., also enforce the length of the 
subharmonic wavevectors to reduce from unity to $|{\bf k}_{Si}|=\sqrt{5/6}\simeq 0.91$.
It can be shown that the fundamental modes contributing to the 
$\sqrt{3}\times \sqrt{3}$ superlattice enter Eq.~\ref{super} with 
 equal spatial phases $\phi_{Hi}=\phi_{Si}=0$. 
The equality $\phi_{H1}=\phi_{H2}=\phi_{H3}=0$ follows
from the mutual resonance among the ${\bf k}_{H}$-modes, while the cross-coupling between 
${\bf k}_S$- and ${\bf k}_H$-modes 
enforces $\phi_{S1}=\phi_{S2}=\phi_{S3}$. The remaining condition $\phi_{Si}=0$ is more subtle 
and established only at quintic order in the 
associated amplitude equation. With the reasoning outlined in Ref.\cite{muller93} one 
obtains either $\phi_{S1}+\phi_{S2}+\phi_{S3}=0$ or $\pi/2$. Since the former (latter)
case is associated with a 6-fold (3-fold) symmetry, we conclude from Fig.\ref{fig4}e that 
$\phi_{Si}=0$ applies in our system. 

If the drive amplitude is rised into phase region $V$ the long 
wavelength modulation of the $\sqrt{3}\times \sqrt{3}$ superlattice disappears and
the pure hexagonal state as depicted in Fig.~\ref{fig4}g survives.
Simultaneously the $\Omega/2$-component in the temporal power spectrum 
of the surface oscillation dies out.  
%
%**************************************************************** 
%*                                                              * 
%*      Summary                                                 * 
%*                                                              * 
%**************************************************************** 
 
To summarize: The present letter reports a Faraday experiment in a very thin 
fluid layer. Drive frequency and filling depth are adjusted
such that waves with subharmonic and harmonic time dependence become simultaneously
unstable. The different time dependencies imply distinct wavelengths, but they are also
responsible for different nonlinear pattern selection mechanisms, which favor 
either squares or hexagons.  In a series of 
bifurcations the transition from one symmetry to the other takes place in a
surprisingly coherent manner running through a 
displacive $\sqrt{2} \times  \sqrt{2}$  and a $\sqrt{3} \times  \sqrt{3}$  superlattice.

A comparison to crystallographic phase transitions is in order. Hexagonal structures on one side and square (in $2D$)
or cubic (in 3D) structures on the other side are incompatible since there is no group-subgroup relation       
connecting the different space groups. Hence such transitions involve a 
reconstruction of the lattice via the formation of lattice defects. The hcp-fcc transition,
occuring e.g. in solid 4He, is a prominent 3D example. Here the relevant defects are
 stacking faults. The same principles hold of course for the present patterns. At the same stage 
along the phase sequence from $II$ to $IV$ there has to be a reconstructive transition involving defects.
 In fact the pattern of Fig.~\ref{fig4}c taken just at the $III-IV$- boundary shows 
such as line defect (running in the vertical direction in the right half of the picture). 
In contrast to the hcp-fcc transition, the transition between the two incompatible symmetries of our system does not
occur in a single step, but involves two intermediate phases.
Transition of the type $II-III$ and $V-IV$ are also known in $2D$ crystallography. The transition
from a simple hexagonal lattice to a $\sqrt{3} \times  \sqrt{3}$  superstructure has been observed 
for
instance in monolayers of $C_2F_5Cl$ adsorbed on graphite  \cite{Fassbender95}. The displacive 
transition $II-III$ is 
analogous to the reconstruction of the (100) surface of Wolfram crystals Ref.~\cite{wolfram90}. 
Here the surface 
atoms are displaced in exactly the same way as the elevation maxima of the surface profile in the 
present study.
In terms of 2D space groups the transition is from $p4$ to $p2mg$ implying a doubling of the unit 
cell. 
The rectangular $p2mg$ symmetry calls for a rectangular metric, that is for different lattice 
parameters 
along two orthogonal directions but this has not observed in our experiment, presumably because 
of the 
insufficient resolution.

%**************************************************************** 
%*                                                              * 
%*      Acknowledgements                                        * 
%*                                                              * 
%**************************************************************** 

{\it Acknowledgements\/} ---  We thank J.~Albers for his support. 
This work is supported by the  Deutsche 
Forschungsgemeinschaft.

%**************************************************************** 
%*                                                              * 
%*      list of references                                      * 
%*                                                              * 
%**************************************************************** 
  
%\newpage

%**************************************************************** 
%*                                                              * 
%*      Figures                                                 * 
%*                                                              * 
%**************************************************************** 

%********************************************************************
%\newpage
%************************************************************************

 %********************************************************************
%\newpage
%************************************************************************

%********************************************************************
%\newpage
%************************************************************************
%
%
%*******************************************************************

%************************************************************************
\end{document}